\documentclass[a4paper,onecolumn,11pt,unpublished]{quantumarticle}
\pdfoutput=1
\usepackage[utf8]{inputenc}
\usepackage[english]{babel}
\usepackage[T1]{fontenc}
\usepackage{amsmath}
\usepackage{amssymb}
\usepackage{mathtools}
\usepackage{booktabs}
\usepackage{graphicx}
\usepackage[numbers]{natbib}
\usepackage{hyperref}

\newcommand{\ket}[1]{\lvert #1 \rangle}

\newcommand{\GF}{\mathrm{GF}}
\newcommand{\Fii}{\mathbb{F}_2}
\newcommand{\Id}{\mathrm{I}}
\newcommand{\pL}{p_{\mathrm{L}}}
\newcommand{\Xerr}{\texttt{X\_ERROR}}
\DeclareMathOperator{\wt}{wt}
\begin{document}

\title{Biased-Noise Quantum Reed--Solomon Codes and a Tornado Concatenation
for Cat Qubits}

\author{Cheng-You Ho$^{*}$}
\author{Daniel Wang$^{*}$}
\author{Simba Shi$^{*}$}
\author{Justin Luo$^{*}$}
\author{Henry Ng$^{*}$}
\affiliation{Yale University, New Haven, Connecticut 06520, USA}

\begin{abstract}
Dissipative cat qubits exponentially suppress one Pauli error channel with the
mean photon number, leaving the conjugate bit-flip error as the dominant
failure mode. This strong noise bias makes the full machinery of general
quantum error correction unnecessary: a code need only protect against a single
error type, and any classical linear code can be promoted to a Clifford
stabilizer code that does exactly this. We use this observation to build a
\emph{bit-flip-only} quantum Reed--Solomon (RS) code. Starting from the
maximum-distance-separable RS code $[7,3,5]$ over $\GF(2^3)$, we binary-expand it
to the linear code $[21,9,6]$ over $\GF(2)$ and realize it as a
$[[21,9,d_X=6,d_Z=1]]$ bit-flip code whose stabilizers are products of $Z$
operators. Because no phase-flip correction is attempted, the construction
discards the redundancy that standard quantum RS codes spend on
correcting $Z$ errors---which a strongly biased cat qubit renders
unnecessary---and yields a shallow Clifford circuit that samples directly in
Stim. Errors are decoded by an optimal bounded-distance syndrome-lookup table. We then introduce a
\emph{Tornado} architecture: a two-layer concatenation that wraps every position
of the outer RS code in an inner distance-three repetition code, yielding a
$[[63,9,18]]$ code decoded by a two-stage inner majority vote and outer lookup
decoder. Monte-Carlo simulation shows that at a physical bit-flip rate $p=0.1$
the Tornado code reaches a logical error rate $\pL \approx 5.3\times10^{-3}$,
below both parent codes, and that its logical error rate scales as $\pL \propto
p^{6}$ at low $p$, in contrast to $p^{2}$ for the repetition code and $p^{3}$
for the standalone RS code. We give the exact construction, the error and
circuit model, an asymptotic scaling analysis, and an honest account of the
overhead cost and single-shot assumptions.
\end{abstract}

\maketitle
\renewcommand{\thefootnote}{}\footnotetext{$^{*}$These authors contributed equally to this work.}\renewcommand{\thefootnote}{\arabic{footnote}}

\section{Introduction}
\label{sec:intro}

Fault-tolerant quantum computing is expensive largely because a generic qubit
suffers two independent kinds of error---bit flips ($X$) and phase flips
($Z$)---and a general-purpose code such as the surface code must spend qubits
suppressing both. Dissipative cat qubits change this accounting. A cat qubit
encodes information in the phase space of a driven-dissipative bosonic mode, and
as the mean photon number grows one of its two error channels is suppressed
\emph{exponentially} while the other grows only polynomially
\cite{mirrahimi2014dynamically,guillaud2019repetition,lescanne2020exponential}.
The result is a qubit with a tunable, and in practice enormous, noise bias:
error rates asymmetric by factors of $10^{5}$ or more have been
demonstrated~\cite{lescanne2020exponential,putterman2025hardware}. Exploiting
such an asymmetry to lower the cost of fault tolerance is a well-established
idea~\cite{aliferis2008fault}. In this
regime one error type is so rare that it can be neglected, and the code above the
cat qubit needs to correct only the dominant one.\footnote{%
In the standard cat-qubit convention the exponentially suppressed error is the
\emph{bit} flip and the dominant residual error is the \emph{phase} flip
\cite{lescanne2020exponential,guillaud2019repetition}; a repetition code that
corrects phase flips is then concatenated on top
\cite{guillaud2019repetition,putterman2025hardware}. The two descriptions are
related by a transversal Hadamard, which exchanges $X\leftrightarrow Z$ and
$XX\leftrightarrow ZZ$ stabilizers. Throughout this paper we adopt the frame used
by our simulation code, in which the single dominant error is labelled $X$ (a
``bit flip'') and is detected by $Z$-type stabilizers. Every statement below
holds verbatim in the phase-flip frame after this relabelling.}

This is a strong simplification, because a code that must correct only one Pauli
error type is essentially a \emph{classical} code. Any binary linear code with
parity-check matrix $H$ becomes a quantum code for the dominant error by
promoting each classical parity check---each row of $H$---to a multi-qubit
$Z$-type stabilizer, so that the classical syndrome and the quantum syndrome
coincide. The encoded states are the classical codewords, the logical operators
are inherited from the classical code, and---crucially---every gate involved
(state preparation, the CNOT-based encoder, stabilizer measurement, and readout)
is Clifford. Such circuits can be simulated exactly and at scale with the
stabilizer simulator Stim~\cite{gidney2021stim}.

This opens the door to high-rate classical codes that would be awkward or
impossible to use as fully quantum ($X$-and-$Z$) codes. Reed--Solomon (RS)
codes~\cite{reed1960polynomial} are the canonical example: they are
maximum-distance separable (MDS), meaning they achieve the largest possible
minimum distance for their rate, and they are ubiquitous in classical storage and
communication. A quantum version of the RS code already exists---the
Grassl--Beth construction~\cite{grassl1999quantum}---built around a discrete
cyclic Fourier transform over the finite field. That transform is a
$\GF(2)$-linear map, so after binary expansion it is an invertible matrix over
$\GF(2)$ that can be realized with CNOT gates; like every stabilizer code, the
Grassl--Beth code therefore admits a Clifford encoder and is in principle
Stim-simulable. However, it is a full CSS code that spends its power correcting both $X$ and
$Z$ errors. Cat qubits render this unnecessary.

In this work we take the opposite, deliberately minimal route. Our contributions
are:
\begin{enumerate}
\item \textbf{A biased-noise quantum RS code.} We construct the RS code $[7,3,5]$
  over $\GF(2^3)$, binary-expand it to $[21,9,6]$ over $\GF(2)$, put it in
  systematic form, and realize it as the bit-flip stabilizer code
  $[[21,9,d_X=6,d_Z=1]]$ with an explicit CNOT encoder and $Z$-type checks
  (Sec.~\ref{sec:qrs}). The code corrects only the dominant $X$ error, so its
  encoder is a shallow CNOT circuit of $Z$-type checks alone, with no
  machinery devoted to phase-flip correction.
\item \textbf{An optimal lookup decoder.} Because the code is small, we decode
  with a maximum-likelihood bounded-distance syndrome table built from the
  systematic parity-check matrix $H=[P^{\mathsf T}\mid \Id]$
  (Sec.~\ref{sec:decoder}).
\item \textbf{The Tornado concatenation.} Inspired by classical Tornado
  codes~\cite{luby2001efficient}, we wrap every position of the outer RS code in
  an inner distance-three repetition code, producing a $[[63,9,18]]$ concatenated
  code decoded in two stages---inner majority vote feeding the outer RS lookup
  (Sec.~\ref{sec:tornado}).
\item \textbf{Benchmarks and scaling.} We compare the repetition, RS, and
  Tornado codes over three orders of magnitude in physical error rate, establish
  the $\pL\propto p^{6}$ low-error scaling of the Tornado code, and give an honest
  discussion of its overhead and of the single-shot noise model
  (Secs.~\ref{sec:results}--\ref{sec:discussion}).
\end{enumerate}

\section{Biased noise and the bit-flip-only code model}
\label{sec:model}

\paragraph{Noise bias.}
We work in the strong-bias limit in which the phase-conjugate error of the cat
qubit is completely suppressed, so that the only error is a bit flip $X$ that
occurs independently on each physical qubit with probability $p$. This is the
idealization used throughout the sponsor challenge and in the cat-qubit
repetition-code literature~\cite{guillaud2019repetition,gouzien2023performance};
it is the leading-order description whenever the bias $\eta = p_{\text{dominant}}
/ p_{\text{suppressed}}$ is large, and it is the regime in which a
single-error-type code is the right tool.

\paragraph{Classical linear codes as quantum bit-flip codes.}
Let $C\subseteq\Fii^{n}$ be a binary linear code with generator matrix $G$
($k\times n$) and parity-check matrix $H$ ($(n-k)\times n$), so that $H
G^{\mathsf T}=0$. We associate one physical qubit with each of the $n$ bits and
define, in the stabilizer formalism~\cite{gottesman1997stabilizer}, the
stabilizer group
\begin{equation}
  \mathcal{S} = \Big\langle\, S_r = \prod_{q:\,H_{rq}=1} Z_q \;:\; r=1,\dots,n-k
  \,\Big\rangle .
  \label{eq:stab}
\end{equation}
All generators are products of $Z$ operators, hence mutually commuting and
Clifford. The codespace is the $+1$ eigenspace of $\mathcal{S}$; its logical
computational states $\ket{c}$ are labelled by the classical codewords $c\in C$,
and a bit-flip pattern $e\in\Fii^{n}$ maps the state $\ket{c}$ to $\ket{c\oplus
e}$. Measuring the stabilizers~\eqref{eq:stab} returns precisely the classical
syndrome
\begin{equation}
  \sigma = H e^{\mathsf T} \in \Fii^{\,n-k},
  \label{eq:syndrome}
\end{equation}
so classical and quantum decoding are identical. In the language of CSS
codes~\cite{calderbank1996good,steane1996error}, this is the degenerate CSS
construction $\mathrm{CSS}(C,\Fii^{n})$ that uses a single classical code to
correct $X$ errors and makes no attempt to correct $Z$ errors. Its quantum
parameters are
\begin{equation}
  [[\,n,\;k,\;d_X = d(C),\;d_Z = 1\,]],
  \label{eq:params}
\end{equation}
where $d(C)$ is the classical minimum distance: a weight-$1$ $Z$ operator is
already a logical operator ($d_Z=1$), which is harmless precisely because the cat
qubit suppresses that error exponentially. Equation~\eqref{eq:params} is the
natural asymmetric-distance generalization of the cat repetition code
$[[n,1,d_X=n,d_Z=1]]$~\cite{guillaud2019repetition}.

\paragraph{Circuit and error model.}
Every logical-error rate reported here is obtained under the following
single-shot memory model, matching our Stim implementation:
(i) all data qubits are reset and the CNOT encoder prepares the logical
all-zero codeword $\ket{0_{\mathrm L}}$;
(ii) each data qubit independently undergoes $X$ with probability $p$
(a single \Xerr$(p)$ location);
(iii) the $Z$-type stabilizers are measured once with a noiseless ancilla,
yielding the syndrome~\eqref{eq:syndrome};
(iv) the data qubits are read out in the $Z$ basis and the $k$ message qubits are
declared as logical observables.
The decoder then applies the correction inferred from the syndrome and a logical
failure is recorded whenever any logical observable is left flipped. This model
isolates the code's combinatorial error-correcting power; measurement and gate
noise and repeated syndrome rounds are deliberately excluded and are discussed as
limitations in Sec.~\ref{sec:discussion}.

\section{Biased-noise quantum Reed--Solomon construction}
\label{sec:qrs}

\subsection{The finite-field engine}
We work over the field $\GF(2^3)=\Fii[x]/(x^3+x+1)$ with primitive polynomial
$x^3+x+1$ (\texttt{0b1011}). Field elements are $3$-bit strings interpreted as
polynomials of degree $<3$; addition is bitwise \textsc{xor} and multiplication
is polynomial multiplication modulo $x^3+x+1$. The element $\alpha=2$ (i.e.\
$x$) is primitive: its powers enumerate all seven nonzero elements,
\begin{equation}
  (\alpha^{0},\dots,\alpha^{6}) = (1,2,4,3,6,7,5),
  \label{eq:evalpts}
\end{equation}
which we use as the evaluation points of the code.

\subsection{Reed--Solomon over $\GF(8)$}
The RS code encodes a message of $k_{\mathrm s}=3$ field symbols
$(m_0,m_1,m_2)$ as the evaluations of the polynomial $m(y)=m_0+m_1 y+m_2 y^2$ at
the $n_{\mathrm s}=7$ points~\eqref{eq:evalpts}:
\begin{equation}
  c_j = \sum_{i=0}^{2} m_i\,\alpha^{\,ij}, \qquad j=0,\dots,6 .
\end{equation}
Equivalently $c = m\,G_{\mathrm{sym}}$ with the $3\times7$ symbol generator
\begin{equation}
  G_{\mathrm{sym}} =
  \begin{pmatrix}
    1 & 1 & 1 & 1 & 1 & 1 & 1\\
    1 & 2 & 4 & 3 & 6 & 7 & 5\\
    1 & 4 & 6 & 5 & 2 & 3 & 7
  \end{pmatrix},
  \label{eq:gsym}
\end{equation}
whose rows are the componentwise powers $(\alpha^{ij})_j$. This is a $[7,3,5]$
code over $\GF(8)$: being an evaluation (generalized Reed--Solomon) code it is
maximum-distance separable, so its symbol distance meets the Singleton bound,
$d_{\mathrm s}=n_{\mathrm s}-k_{\mathrm s}+1=5$, and it corrects any
$t=\lfloor(d_{\mathrm s}-1)/2\rfloor=2$ symbol errors.

\subsection{Binary expansion and systematic form}
\begin{figure}[t]
  \centering
  \includegraphics{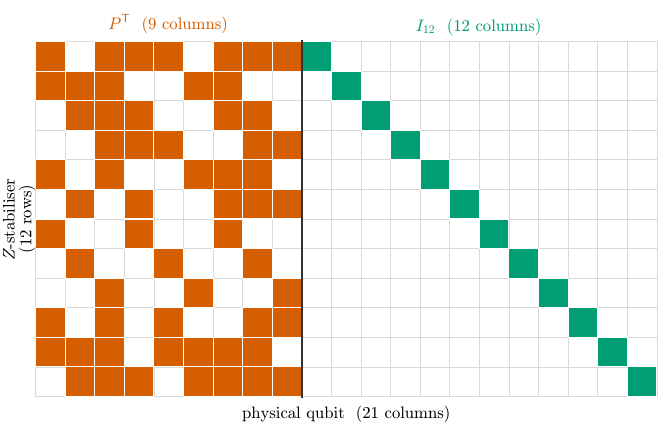}
  \caption{The systematic parity-check matrix $H=[P^{\mathsf T}\mid\Id_{12}]$ of
    the binary-expanded quantum Reed--Solomon code $[[21,9,6]]$. Each of the $12$
    rows is a $Z$-type stabilizer generator; each of the $21$ columns is a
    physical qubit. Filled cells mark a $1$. The left block $P^{\mathsf T}$
    (vermillion) couples parity checks to the $9$ message qubits; the right block
    is the $12\times12$ identity (green) acting on the parity qubits. $H$ has $72$
    nonzero entries.}
  \label{fig:paritycheck}
\end{figure}

Quantum hardware acts on bits, not field symbols, so we expand
$G_{\mathrm{sym}}$ to $\GF(2)$ by replacing each symbol operation with its action
on the underlying $3$-bit strings. Encoding the $9$ basis messages produces a
$9\times21$ binary generator matrix $G_{\mathrm{bin}}$. Gaussian elimination over
$\GF(2)$, followed by a column permutation that collects the pivots, brings it to
systematic form
\begin{equation}
  G_{\mathrm s} = [\,\Id_{9}\mid P\,], \qquad P\in\Fii^{\,9\times12},
\end{equation}
from which the systematic parity-check matrix follows immediately,
\begin{equation}
  H = [\,P^{\mathsf T}\mid \Id_{12}\,] \in \Fii^{\,12\times21},
  \qquad H G_{\mathrm s}^{\mathsf T}=0 .
  \label{eq:H}
\end{equation}
A brute-force minimum-weight search over all $2^{9}-1$ nonzero codewords gives a
binary minimum distance of $6$; the binary image of the RS code is therefore the
linear code $[21,9,6]$, corrects $t=2$ bit errors, and as a quantum bit-flip code
has parameters
\begin{equation}
  [[\,21,\,9,\,d_X=6,\,d_Z=1\,]].
\end{equation}
The $12$ rows of $H$ are the supports of the $12$ $Z$-type stabilizers; its
structure is shown in Fig.~\ref{fig:paritycheck}. The systematic layout separates
the $21$ physical qubits into $9$ message qubits and $12$ parity qubits and
directly dictates the encoder.

\subsection{The encoder and stabilizer circuit}
The systematic generator prepares the logical all-zero codeword from the physical
all-zero state using only CNOTs: for every $1$ in $P$ we entangle the
corresponding message and parity qubit,
\begin{equation}
  \text{if } P_{ij}=1:\quad \mathrm{CX}(\,\text{message } i \to
  \text{parity } 9+j\,).
\end{equation}
Syndrome extraction measures each stabilizer $S_r$ of Eq.~\eqref{eq:stab} by
copying the parities of its support onto a fresh ancilla with CNOTs and measuring
the ancilla in the $Z$ basis, contributing one detector bit per row of $H$.
Finally the $9$ message qubits are read out and declared as logical observables.
Every operation is Clifford; the circuit compiles and samples directly in
Stim~\cite{gidney2021stim}. This should be contrasted with the Grassl--Beth
quantum RS code~\cite{grassl1999quantum}, a full CSS code that spends part of its structure on
$Z$-error correction that a cat qubit does not need. Our encoder omits that
$Z$-correcting machinery entirely, leaving the shallow CNOT circuit above.

\subsection{Maximum-likelihood lookup decoder}
\label{sec:decoder}
For a code this small, optimal decoding is a table lookup. We precompute, for
every attainable syndrome, the minimum-weight error consistent with it. Iterating
over error patterns of increasing weight $w=0,1,2,3$ and keying by the syndrome
$H e^{\mathsf T}$, the first pattern found for each syndrome is a minimum-weight
coset leader:
\begin{equation}
  \hat e(\sigma) = \arg\min_{e:\,He^{\mathsf T}=\sigma} \wt(e).
\end{equation}
Enumerating up to weight $3$ yields a table with $1359$ distinct syndromes out of
the $2^{12}=4096$ possible. All weight-$\le2$ errors are corrected exactly (the
guaranteed bounded-distance radius $t=2$), and the weight-$3$ leaders fill in
additional syndromes so that some weight-$3$ errors are corrected as well. A
logical failure is declared when the residual error $e\oplus\hat e$ is a nonzero
codeword, i.e.\ when the decoder returns to a wrong codeword or the syndrome lies
outside the table. We verified that this classical decoder, applied directly to random error
patterns, agrees with the same decoder applied to the detector samples of the
full Stim circuit to within Monte-Carlo error (e.g.\ $\pL=0.182$ vs.\ $0.181$ at
$p=0.1$, consistent with Fig.~\ref{fig:benchmark}), as it must for a purely
classical $X$-error channel.

\section{The Tornado concatenation}
\label{sec:tornado}

\subsection{Motivation}
The RS code is high-rate but fragile: with $d_X=6$ it tolerates only two bit flips
among $21$ qubits, so at large $p$ its logical error rate is actually
\emph{worse} than a simple repetition code (Fig.~\ref{fig:benchmark}). A
repetition code is the opposite---extremely robust but wasteful of qubits.
Classical Tornado codes~\cite{luby2001efficient} resolve exactly this tension by
layering sparse graph codes with an outer high-rate code so that each layer cleans
up the errors that slip past the previous one. We adopt a deliberately simple,
two-layer instance of this idea suited to biased-noise qubits.

\subsection{Architecture}
\begin{figure*}[t]
  \centering
  \includegraphics{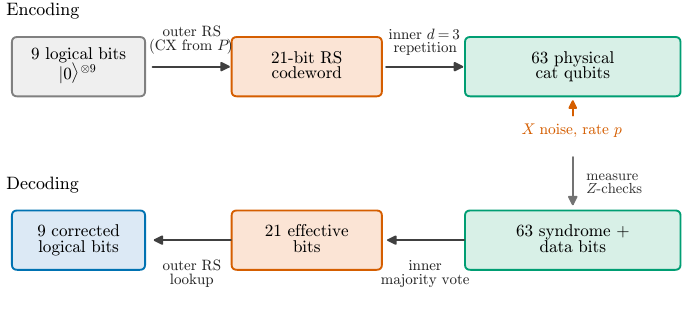}
  \caption{The Tornado architecture. \emph{Encoding} (top): nine logical bits are
    encoded by the outer Reed--Solomon code into a $21$-bit codeword using the
    CNOT encoder read off from $P$; every one of those $21$ positions is then
    wrapped in an inner distance-three repetition code, giving $63$ physical cat
    qubits exposed to bit-flip noise at rate $p$. \emph{Decoding} (bottom): the
    $63$ measured qubits are first collapsed by an inner majority vote (one vote
    per repetition block) to $21$ effective bits, which are then decoded by the
    outer Reed--Solomon syndrome-lookup table to recover the nine logical bits.}
  \label{fig:architecture}
\end{figure*}

The Tornado code concatenates an \emph{inner} distance-three repetition code
inside the \emph{outer} RS code, as sketched in Fig.~\ref{fig:architecture}. Each
of the $n=21$ positions of an RS codeword is itself encoded into a block of
$d_{\mathrm{in}}=3$ physical cat qubits by a repetition code. The composite code
therefore uses
\begin{equation}
  N = n\cdot d_{\mathrm{in}} = 21\times3 = 63
\end{equation}
physical qubits to protect $k=9$ logical qubits. Concatenation multiplies
distances, so the $X$-distance of the composite code is
\begin{equation}
  d_X = d_{\mathrm{in}}\cdot d(C) = 3\times6 = 18,
\end{equation}
while $d_Z$ remains $1$; the Tornado code is $[[63,9,18]]$. Its rate is $9/63=1/7$.

\subsection{Two-stage decoder}
Decoding mirrors the encoding, from the inside out. For each of the $21$
repetition blocks the \emph{inner} decoder takes a hard majority vote over its
$d_{\mathrm{in}}=3$ qubits, producing a single effective bit. A block is decoded
incorrectly exactly when at least two of its three qubits flipped, which happens
with the effective error probability
\begin{equation}
  q(p) = \binom{3}{2}p^{2}(1-p) + p^{3} = 3p^{2}-2p^{3} \approx 3p^{2}.
  \label{eq:qeff}
\end{equation}
The $21$ effective bits are then handed to the \emph{outer} decoder, which is
exactly the RS syndrome-lookup table of Sec.~\ref{sec:decoder} applied with error
rate $q(p)$ in place of $p$. Because the noise is a classical $X$ channel and both
layers are CSS, this two-stage hard-decision procedure is equivalent to a
stabilizer simulation of the full $63$-qubit circuit and we evaluate it directly
by Monte-Carlo sampling.

\section{Numerical results}
\label{sec:results}

\begin{figure}[t]
  \centering
  \includegraphics{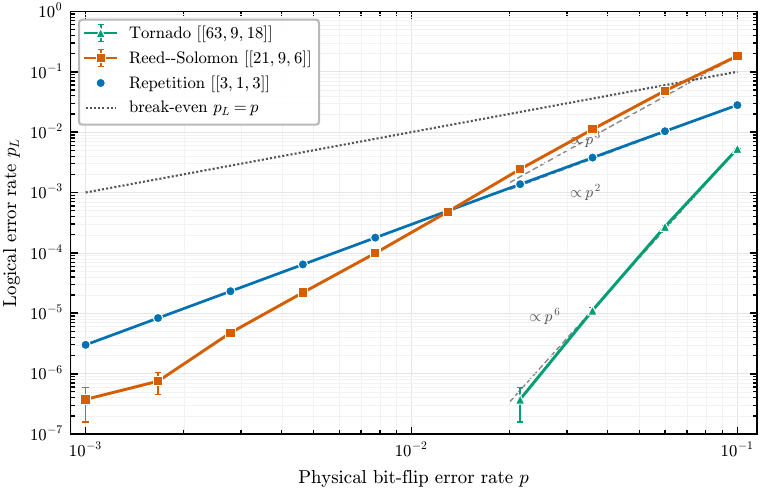}
  \caption{Logical error rate $\pL$ versus physical bit-flip rate $p$ for the
    three codes, on log--log axes. The Tornado code (green triangles) lies below
    both parent codes wherever it is resolvable and falls off far more steeply.
    Dashed guides show the asymptotic slopes $\pL\propto p^{2}$ (repetition),
    $p^{3}$ (Reed--Solomon), and $p^{6}$ (Tornado); the dotted line is the
    uncoded break-even $\pL=p$. Error bars are $1\sigma$ Poisson; where the
    Tornado rate falls below the sampling floor the markers are omitted.}
  \label{fig:benchmark}
\end{figure}

Figure~\ref{fig:benchmark} is the central result: the logical error rate $\pL$ of
the repetition code $[[3,1,3]]$, the standalone RS code $[[21,9,6]]$, and the
Tornado code $[[63,9,18]]$ as a function of the physical bit-flip rate $p$, over
$p\in[10^{-3},10^{-1}]$. The repetition curve is the exact majority-vote
expression $3p^2-2p^3$; the RS and Tornado curves are vectorized Monte-Carlo runs
of the decoders of Secs.~\ref{sec:decoder} and~\ref{sec:tornado}, with up to
$8\times10^{6}$ shots per point and Poisson error bars. Representative values are
collected in Table~\ref{tab:results}.

\begin{table}[t]
  \centering
  \caption{Logical error rate $\pL$ at three physical error rates, and the
    factor by which the Tornado code improves on each parent. Monte-Carlo values;
    the repetition entry is exact.}
  \label{tab:results}
  \begin{tabular}{lccc}
    \toprule
    & $p=0.036$ & $p=0.06$ & $p=0.1$\\
    \midrule
    Repetition $[[3,1,3]]$   & $3.8\times10^{-3}$ & $1.0\times10^{-2}$ & $2.8\times10^{-2}$\\
    Reed--Solomon $[[21,9,6]]$ & $1.1\times10^{-2}$ & $4.8\times10^{-2}$ & $1.8\times10^{-1}$\\
    Tornado $[[63,9,18]]$    & $1.1\times10^{-5}$ & $2.7\times10^{-4}$ & $5.3\times10^{-3}$\\
    \midrule
    gain vs.\ repetition     & $340\times$ & $38\times$ & $5.3\times$\\
    gain vs.\ Reed--Solomon  & $990\times$ & $180\times$ & $34\times$\\
    \bottomrule
  \end{tabular}
\end{table}

Three features stand out. First, the RS and repetition curves cross near
$p\approx0.013$: below it the high-rate RS code wins, above it the robust
repetition code wins, quantifying the density-versus-robustness trade-off.
Second, the Tornado code is below both parents throughout the resolvable range;
at $p=0.1$ it reaches $\pL\approx5.3\times10^{-3}$, a factor of $5.3$ below the
repetition code and $34$ below the standalone RS code. Third, and most
importantly, the Tornado advantage \emph{grows} as $p$ decreases because its
curve is steeper: by $p\approx0.036$ the improvement is already $340\times$ over
repetition and $990\times$ over RS (Table~\ref{tab:results}). The concatenation
does not merely shift the curve down, it bends it.

\section{Asymptotic scaling}
\label{sec:scaling}

The slopes in Fig.~\ref{fig:benchmark} follow from counting the minimum-weight
error that defeats each decoder. A code that corrects $t$ errors first fails at
weight $t+1$, so its logical error rate scales as $\pL\propto p^{\,t+1}$ at small
$p$.

\paragraph{Repetition ($d=3$).} Majority vote fails at two flips out of three, so
$\pL^{\mathrm{rep}}\approx 3p^{2}$: slope $2$.

\paragraph{Reed--Solomon $[[21,9,6]]$.} The bounded-distance decoder corrects
$t=2$ bit flips and first fails at weight $3$, so $\pL^{\mathrm{RS}}\sim A_3\,
p^{3}$: slope $3$. (The weight-$3$ coset leaders in the table lower the prefactor
$A_3$ but do not change the exponent.)

\paragraph{Tornado $[[63,9,18]]$.} A logical failure requires the outer RS decoder
to fail, i.e.\ at least $t+1=3$ of the $21$ effective bits to be wrong; each
effective bit is wrong with probability $q(p)\approx3p^{2}$ from
Eq.~\eqref{eq:qeff}. Hence
\begin{equation}
  \pL^{\mathrm{tor}} \;\sim\; \binom{21}{3}\,q(p)^{3}
  \;\approx\; \binom{21}{3}\,(3p^{2})^{3}
  \;=\; 3.6\times10^{4}\,p^{6},
  \label{eq:p6}
\end{equation}
a slope of $6$. A least-squares fit to the simulated Tornado data over
$p\in[0.03,0.06]$ gives a local slope of $6.3$, consistent with
Eq.~\eqref{eq:p6}; the small excess over $6$ and the smaller measured prefactor
both reflect the weight-$3$ coset leaders, which correct a fraction of the
three-block failures. More generally, concatenating an inner code whose failure
probability scales as $p^{a}$ with an outer code that first fails at $b$ block
errors yields an exponent $a\,b$; here $a=2$ (majority vote) and $b=3$ (RS
correcting two), giving $6$.

It is worth being precise about what this exponent is and is not. The code
distance $d_X=18$ would, under an optimal decoder, permit correcting up to
$\lfloor(18-1)/2\rfloor=8$ errors and yield a slope of $\lceil 18/2\rceil=9$. Our
hard-decision, two-stage decoder does not reach that: it fails already at $6$
physical flips (two in each of three blocks), so its exponent is $6$, not $9$. The
gap is the price of decoding the two layers independently and with hard decisions;
a soft-decision decoder acting on the full $63$-qubit code could in principle
recover the missing suppression. We regard the $p^{6}$ scaling as the honest,
achieved performance of the simple decoder we implement, with headroom identified
for future work.

\section{Discussion and limitations}
\label{sec:discussion}

\paragraph{Overhead is the cost of the steep slope.}
The Tornado code's low logical error rate is bought with qubits, and it is
important to state this plainly. Table~\ref{tab:overhead} lists the rates: the
Tornado code has the \emph{lowest} rate of the three, $1/7$, below both the
standalone RS code ($3/7$) and even the $d=3$ repetition code ($1/3$). The
Tornado code is not a free improvement over its parents at fixed qubit budget; it
trades rate for a much steeper error suppression. The fair comparison is at fixed
\emph{distance}: a pure repetition code of distance $18$ would need $18$ physical
qubits per logical qubit (rate $1/18\approx0.056$), whereas the Tornado code
achieves the same distance-$18$ protection at rate $1/7\approx0.143$, a
$2.6\times$ improvement in encoding rate. This is the concrete sense in which
folding a high-rate algebraic outer code into the concatenation is worthwhile:
for a target distance it is far more qubit-efficient than brute-force repetition,
which is the same lesson that motivates high-rate outer codes in the cat-qubit
architectures of Refs.~\cite{gouzien2023performance,ruiz2025ldpccat,
putterman2025hardware}.

\begin{table}[t]
  \centering
  \caption{Resource comparison. ``Overhead'' is physical qubits per logical
    qubit. The Tornado code trades the lowest rate for the largest distance and
    steepest error suppression.}
  \label{tab:overhead}
  \begin{tabular}{lcccc}
    \toprule
    Code & $[[n,k,d_X]]$ & rate $k/n$ & overhead & $\pL\propto$\\
    \midrule
    Repetition   & $[[3,1,3]]$   & $0.33$ & $3$    & $p^{2}$\\
    Reed--Solomon & $[[21,9,6]]$  & $0.43$ & $2.3$  & $p^{3}$\\
    Tornado      & $[[63,9,18]]$ & $0.14$ & $7$    & $p^{6}$\\
    \bottomrule
  \end{tabular}
\end{table}

\paragraph{Single-shot, code-capacity noise.}
Our results are code-capacity results: a single round of $X$ noise followed by
\emph{noiseless} syndrome extraction and readout. Real devices have faulty gates
and measurements and require repeated syndrome rounds, under which an outer code
with only $d_Z=1$ and a static, single-shot decoder is not by itself
fault-tolerant. Establishing a circuit-level threshold would require modelling
CNOT and measurement noise, decoding over multiple rounds, and accounting for the
residual phase-flip rate of the physical cat qubits (which sets the $d_Z=1$
vulnerability). We view the present construction as a code-capacity proof of
concept for using high-rate classical codes on biased-noise qubits, not as a
fault-tolerance claim.

\paragraph{Decoder and code-family scaling.}
The lookup decoder is optimal but its table grows exponentially in the number of
correctable errors, so it does not scale to large codes; a syndrome-based
algebraic RS decoder or belief propagation would be needed there. Likewise, the
three codes benchmarked here are single instances, not a family, so
Fig.~\ref{fig:benchmark} exhibits finite-size scaling rather than a threshold. The
construction is nonetheless parametric: taking RS$[15,9]$ over $\GF(2^4)$
(primitive polynomial $x^4+x+1$) binary-expands to a $[60,36]$ code with symbol
distance $7$ and binary distance at least $7$, and the same Tornado wrapping
applies to any of these. Mapping out a genuine threshold for a scalable Tornado
family---and comparing against tailored quantum LDPC codes for biased
noise~\cite{tuckett2019tailoring,roffe2023biastailored,ruiz2025ldpccat}---is the
natural next step.

\paragraph{Relation to prior and concurrent work.}
Our construction sits at the intersection of three lines of work. Against the
Grassl--Beth quantum RS codes~\cite{grassl1999quantum} we trade generality
(correcting both $X$ and $Z$) for an $X$-only construction exactly matched to the
biased cat-qubit channel---a strictly simpler, shallower circuit obtained by
dropping the $Z$-correcting half of the code. Against classical Tornado
codes~\cite{luby2001efficient} we keep only the two-layer
outer-algebraic/inner-simple concatenation philosophy, replacing the LDPC cascade
with a single repetition layer. And within the cat-qubit literature, concatenating
biased qubits with a classical outer code is precisely the strategy behind
repetition-cat architectures~\cite{guillaud2019repetition,gouzien2023performance,
putterman2025hardware} and the LDPC-cat~\cite{ruiz2025ldpccat} and
bias-tailored-LDPC~\cite{roffe2023biastailored} codes; the recent Elevator
codes~\cite{shanahan2026elevator} similarly concatenate an inner repetition layer
with a high-rate outer code under biased noise. Our contribution is a small,
fully explicit, optimally decoded instance built from a maximum-distance-separable
Reed--Solomon outer code, together with a clean account of its scaling and
overhead.

\section{Conclusion}
\label{sec:conclusion}
Exploiting the noise bias of cat qubits, we built a bit-flip-only quantum
Reed--Solomon code---the binary-expanded $[[21,9,6]]$ code---that retains only
the $X$-correcting half of a full CSS quantum RS construction, yielding a
shallow Clifford circuit simulable end-to-end in Stim. Wrapping each of its positions
in a distance-three repetition code yields a $[[63,9,18]]$ Tornado code that, with
a simple two-stage decoder, reaches a logical error rate of $5.3\times10^{-3}$ at
a physical error rate of $0.1$ and suppresses errors as $p^{6}$, far steeper than
either parent code. The construction is a compact demonstration that the extreme
noise asymmetry of bosonic qubits turns the rich toolbox of classical linear codes
into directly usable, Clifford quantum codes, and that concatenating a high-rate
algebraic outer code with a simple inner code is a qubit-efficient route to steep
error suppression. Extending the analysis to circuit-level noise and to a scalable
code family are the clear next steps.

\begin{acknowledgments}
We thank the organizers of MIT iQuHACK 2026 and Alice~\&~Bob for the cat-qubit
challenge that seeded this work, and mentors Shantanu Jha and Diego Polimeni for
helpful discussions. Numerical work used Stim~\cite{gidney2021stim} and
PyMatching~\cite{higgott2022pymatching,higgott2025sparseblossom}.
\end{acknowledgments}

\bibliographystyle{quantum}
\bibliography{references}

\appendix

\section{Reproducibility}
\label{app:repro}
All code, data, and figure scripts are available in the project repository at
\url{https://github.com/Coderrexe/2026-Alice-and-Bob} (commit \texttt{0089ab3}).
The numerical results in this paper were reproduced independently of the original
notebook using Python~3, \texttt{stim}~1.16.0, \texttt{pymatching}~2.4.0,
\texttt{numpy}, and \texttt{scipy}; the original notebook pins \texttt{stim}\,$\sim$\,1.15.
The construction constants---primitive polynomial $x^3+x+1$, primitive element
$\alpha=2$, evaluation points $(1,2,4,3,6,7,5)$, the generator
$G_{\mathrm{sym}}$ of Eq.~\eqref{eq:gsym}, the binary distance $6$, and the
$1359$-entry lookup table---were all verified against a clean re-implementation.
The self-contained module \texttt{figures/rs\_tornado\_lib.py} rebuilds the code,
$H$, and the decoders; \texttt{figures/make\_data.py} regenerates the benchmark
data of Fig.~\ref{fig:benchmark} and Table~\ref{tab:results}; and the
\texttt{figures/fig\_*.py} scripts regenerate every figure.

\section{Explicit code matrices}
\label{app:matrices}
For completeness we give the code in fully explicit form, so that it can be
reconstructed without running any code. The Gaussian elimination of
Sec.~\ref{sec:qrs} yields the systematic generator $G_{\mathrm s}=[\,\Id_9\mid
P\,]$ with the $9\times12$ parity block
\begin{equation}
  P =
  \setlength{\arraycolsep}{3.5pt}
  \begin{pmatrix}
    1&1&0&0&1&0&1&0&0&1&1&0\\
    0&1&1&0&0&1&0&1&0&0&1&1\\
    1&1&1&1&1&0&0&0&1&1&1&1\\
    1&0&1&1&0&1&1&0&0&0&0&1\\
    1&0&0&1&0&0&0&1&0&1&1&0\\
    0&1&0&0&1&0&0&0&1&0&1&1\\
    1&1&1&0&1&1&1&0&0&0&1&1\\
    1&0&1&1&1&1&0&1&0&1&1&1\\
    1&0&0&1&0&1&0&0&1&1&0&1
  \end{pmatrix}.
  \label{eq:Pexplicit}
\end{equation}
The parity-check matrix is $H=[\,P^{\mathsf T}\mid\Id_{12}\,]$
(Eq.~\eqref{eq:H}), the $12$ rows of which are the $Z$-type stabilizer supports;
one verifies $H\,G_{\mathrm s}^{\mathsf T}=0$ over $\GF(2)$. The encoder places a
CNOT from message qubit $i$ to parity qubit $9+j$ for every nonzero entry
$P_{ij}$ of Eq.~\eqref{eq:Pexplicit}, giving $72-12=60$ encoding CNOTs (equal to
the number of ones in $P$; the $12$ ones on the diagonal of the $\Id_{12}$ block
of $H$ are stabilizer self-supports, not gates). A brute-force search over the
$2^{9}-1$ nonzero codewords of $G_{\mathrm s}$ confirms the minimum distance $6$.

\section{Worked field arithmetic}
\label{app:gf}
In $\GF(8)=\Fii[x]/(x^3+x+1)$ the relation $x^3=x+1$ reduces higher powers.
Writing elements as $3$-bit strings (little-endian in the powers of $x$), the
element $2=\texttt{010}=x$ generates the multiplicative group:
$x^{1}=2$, $x^{2}=4$, $x^{3}=x+1=3$, $x^{4}=x^{2}+x=6$, $x^{5}=x^{2}+x+1=7$,
$x^{6}=x^{2}+1=5$, $x^{7}=1$, reproducing the evaluation points of
Eq.~\eqref{eq:evalpts}. Addition is \textsc{xor}: e.g.\ $3\oplus3=(x{+}1)\oplus(x{+}1)=0$.
Multiplication follows the shift-and-reduce rule; e.g.\
$6\cdot6=(x^{2}{+}x)^{2}=x^{4}+x^{2}=x(x{+}1)+x^{2}=x=2$.

\end{document}